# Transversal electric field effect in multilayer graphene nanoribbon

S. Bala kumar and Jing Guo[a]

*Department of Electrical and Computer Engineering, University of Florida, Gainesville, Florida 32608, USA*

Abstract

We study the effect of transversal electric-field (E-field) on the electronic properties of multilayer armchair-graphene-nanoribbon (AGNR). The bandgap in multilayer-AGNRs can be reversibly modulated with the application of E-field. At optimized widths, we obtain a semiconductor (SC) to metallic (M), as well as M-SC transition. The AGNR electronic bands undergo vivid transformations due to the E-field, leading to phenomena such as increase in electron velocity, change in the sign of the electron effective mass, and the formation of linear dispersion with massless Dirac fermions similar to 2D-graphene. These effects are very useful and can be utilized for device applications.

---

[a] Corresponding author. E-mail: guoj@ufl.edu

Graphene, a two-dimensional carbon sheet, has attracted various interesting studies on transport[1-3], magnetic[4-7], and optical[7,8] properties. Graphene is a zero band gap semiconductor with unique linear dispersion around the Brillouin zone corners[9-11]. The absence of band gap renders graphene challenging in the application of electronic and optical devices. One of the methods to increase the band gap of graphene is to form graphene nanoribbon (GNR)[12-15] such that electrons are confined in the transversal direction. Based on the shape of the edges, GNRs can be classified into armchair GNRs (AGNRs) or zigzag GNRs (ZGNRs). AGNRs are semiconducting with the finite band gaps. Recently, it has been shown that the electronic properties of AGNR can be continuously and reversibly modulated by an external transversal electric field (E-field)[16-19].

In this letter, we focus on the effect of transversal E-field on the electronic structure of the multilayer AGNR using nearest neighbor π-orbital tight binding model[9, 20-23]. We find that for the multilayer AGNR, the band gap can be either increased or decreased by optimizing the E-field and AGNR width. Semiconductor to metal transition, as well as metal to semiconductor transition is obtained by varying the E-field. We also show significant transformations of the electronic bands of the multilayer AGNR, which lead to the effects such as increase in electron velocity (GNR generally have velocities lesser than 2D-graphene), change in the sign of the electron effective mass, and formation of linear dispersion with massless Dirac fermions similar to 2D-graphene. The modulations of these properties, plays an important role in electronic and optical device applications.

The inset of Figure 1(b) shows the schematic diagram of AGNR multilayers in transversal E-field. Side gates are used to apply the E-field. The asymmetric electric potential varies linearly across the width of the ribbon. The total number of graphene layers is m. These layers are stacked in AB-stacking alignment. Each layer consists of an AGNR with n dimer rows. In this letter, we label an AGNR with n dimers as an n-AGNR.

First we examine the band gap variation in multilayer AGNR under zero E-field. A widely used nearest neighbor π-orbital tight binding model[9, 20-23] is used in this letter. The Hamiltonian for graphite in AB-stacking is given by

$$h(k) = \begin{bmatrix} 0 & \lambda & 0 & 0 \\ \lambda^* & 0 & \Lambda & 0 \\ 0 & \Lambda^* & 0 & \lambda \\ 0 & 0 & \lambda^* & 0 \end{bmatrix}, \quad (1)$$

,where $\lambda = t_0 \left(1 + \exp(ik \bullet a_1) + \exp(ik \bullet a_2)\right)$, $\Lambda = t_\perp \left(1 + \exp(ik_z \bullet c)\right)$, $a_1 = \left(3/2, +\sqrt{3}/2\right) a_{cc}$, $a_2 = \left(3/2, -\sqrt{3}/2\right) a_{cc}$, and $k = (k_x, k_y)$. The interlayer (intralayer) hopping parameter is $t_\perp$ ($t_0$). The interlayer (intralayer) C-C bond length is c/2 ($a_{cc}$). Along the Γ-K line ($k_x$=0),

$$|\lambda| = t_0 \left|1 + 2\cos\frac{\sqrt{3}k_y a_{cc}}{2}\right| \quad (2a)$$

$$|\Lambda| = 2t_\perp \left|\cos\frac{k_z c}{2}\right| \quad (2b)$$

To form GNR, the graphite is confined in the y- and z-directions. The E-k relation of the AGNR can be obtained analytically by using the zone-folding method. Here we neglect the edge-bond relaxation, which only adds quantitative perturbations to the bandstructure. In AGNR the wave function is zero at the edges. Therefore, $\sin\left((Y + \sqrt{3}a_{cc})k_y\right) = 0$ and $\sin\left((Z + c)k_z\right) = 0$, where $Y = \sqrt{3}a_{cc}(n-1)/2$ and $Z = c(m-1)/2$ are the width and the thickness of the multilayer AGNR, respectively. Solving these equations, we obtain

$$k_y a_{cc} = \frac{2}{\sqrt{3}} \frac{\pi \gamma}{n+1} \Rightarrow |\lambda| = t_0 \left| 1 + 2\cos\frac{\pi\gamma}{n+1} \right| \tag{3a}$$

$$k_z C = 2\frac{\pi\alpha}{m+1} \Rightarrow |\Lambda| = 2t_\perp \left|\cos\frac{\pi\alpha}{m+1}\right| \tag{3b}$$

where γ=1,2,...,n and α=1,2,...,m. The eigenenergy of eq (1), E is such that

$$|E|^2 = |\lambda|^2 + \frac{|\Lambda|^2}{2} \pm \frac{|\Lambda|}{2}\sqrt{4|\lambda|^2 + |\Lambda|^2} \tag{4}$$

The band gap of a multilayer AGNR is given by $E_g = 2|E|_{min}$, where $|E|_{min}$ is the minimum value of |E| as we vary γ and α. From eq (3) and eq (4) it can been shown that $|E|_{min}$ is obtained when $|\lambda| \to 0$ and $|\Lambda| \to 2t_\perp$. These conditions are satisfied when

$$\gamma = \lceil (2n+1)/3 \rceil \Rightarrow |\lambda| = t_0 \left| 1 + 2\cos\frac{\pi\lceil (2n+1)/3 \rceil}{n+1} \right| \equiv \lambda_g \tag{5a}$$

$$\alpha = m \Rightarrow |\Lambda| = 2t_\perp \left|\cos\frac{\pi m}{m+1}\right| \equiv \Lambda_g \tag{5b}$$

Based on Eg. (4) and (5) the expression for the band gap in a multilayer AGNR is such that $E_g^2 = 4\lambda_g^2 + 2\Lambda_g^2 - 2\Lambda_g\sqrt{4\lambda_g^2 + \Lambda_g^2}$. Note that $E_g$ is suppressed when $|\lambda| \to 0$ and $|\Lambda| \to 2t_\perp$.

Referring to eq (5), for any fixed number of layers, m, when the width of the AGNR in each layer, n increases, $|\lambda|$ (and thus $E_g$) oscillates with a period of three, with a suppressing magnitude. When n=3p-1 [p=1,2,3,...], $E_g$=0, and when n≠3p-1, Eg>0. This is shown in Figure

1, which is computed using the nearest neighbor π-orbital TB model with edge bond relaxation. The results show that, just like monolayer AGNRs[13], multilayer AGNRs can be divided into three families: (3p-1)-AGNR, 3p-AGNR, and (3p+1)-AGNR. The band gap varies in the following order: (3p-1)-AGNR > 3p-AGNR > (3p+1)-AGNR; and for each type, the band gap decreases with increasing width. Due to the edge-bond relaxation, the band gap at n=3p-1, is small but not zero. Figure 1 also shows that when the number of layers increases, the band gap is suppressed for all the three families. This is because when m increases the $|\Lambda| \to 2t_\perp$ and thus the $E_g$ is suppressed [refer Eq. 5]. The inset of Figure 1(c) shows that the effect of edge-bond relaxation, which induces a finite band gap in (3p-1)-AGNR, is suppressed when the number of layers is increased.

Next we investigate the effect of transversal E-field in the multilayer AGNR. In Figure 2 we show the band gap variation with varying widths and E-field, for different number of layers. For the monolayer AGNR, in general, the band gap decreases with increasing E-field. This result has been shown in the previous theoretical studies[17-19]. On the other hand, for multilayer AGNRs, the band gap variation under E-field is significantly different from the monolayer AGNRs. This is mainly due to the difference in the charge distribution across the AGNRs, which in turn varies the effective electric potential at the band edges, as thus resulting in different band gap variations. For m>2, we find that a significant increase in band gap is obtained at the optimized widths. For very narrow ribbons, the band gap is decreased with increasing E-field. For very wide ribbons, even though the band gap increases with E-field, the increase is very small. At the optimized width we can obtain a significant increment in the band gap. However, the maximum value of the band gap is around 0.2eV. This maximum value imposes a limitation on the band gap modulation in multilayer AGNR using transversal E-field.

To study the band gap modulation effect in detail, we choose two AGNR's- a (3p-1)-AGNR and a (3p+1)-AGNR. Figure 3a shows the effect of transversal E-field on (3p-1)-AGNR. At zero E-field, the monolayer (3p-1)-AGNR has a small bandgap. However for the multilayer AGNR, i.e. m>2, the (3p-1)-AGNR is metallic (M) with zero bandgap. When E-field is applied, the band gap increases to a maximum value and then decreases back to zero. The bilayer (3p-1)-AGNR exhibits the largest band gap increment. For the (3p+1)-AGNR the variation is shown in Figure 3b. At zero E-field, (3p+1)-AGNR is semiconducting (SC) with a relatively large band gap. With the increase of the field, this maximum value decreases to zero, and then again increases to a large value. In general, for the multilayer (m>2) AGNRs, the (3p-1)-AGNR ((3p+1)-AGNR) shows a M-SC-M (SC-M-SC) transition.

Next we study the effect of transversal E-field on the density of states (DOS) and the band structure of the AGNRs, especially at the transitional points. We choose the bilayer 28-AGNR and trilayer 29-AGNR as examples, since these two AGNRs show a very prominent effect of M-SC transition. Note that the width of the bilayer 28-AGNR (trilayer 29-AGNR) is 3.37nm (3.50nm). Figure 4 (ai) and (bi) show the DOS for the bilayer 28-AGNR and trilayer 29-AGNR, respectively. The "black" regions indicate the band gap where DOS is zero, while the brighter regions indicate higher DOS values. Figure 4ai (Figure 4bi) show that a clear M-SC-M (SC-M-SC) transition is seen at $\varepsilon_1$- $\varepsilon_2$- $\varepsilon_3$ for the bilayer 29-AGNR (trilayer 28-AGNR).

Figure 4aii shows the band structure of the bilayer 29-AGNR, at the three critical points, i.e. $\varepsilon_1=0$, $\varepsilon_2=0.27$V/nm and $\varepsilon_3=0.47$V/nm. When $\varepsilon=\varepsilon_1$, the band gap is almost zero (M-state). As the E-field increases, the band gap increases and the ribbon becomes SC. When the ribbon is SC, the minimum |E| occurs at |k|>0. As a result, the conduction band has a negative curvature at k=0, indicating a negative electron effective mass. As the $\varepsilon$ further increases, the minimum |E|

decreases to zero (M-state) at |k|>0 while the band flattens at k=0. Figure 4(bii) shows the band structure variation for trilayer 28-AGNR. Being a SC, the trilayer 28-AGNR has a finite band gap at $\varepsilon=0$. As the $\varepsilon$ increases, the gradient of E-k around k=0 increases, and the minimum |E| decreases, approaching zero. The increase in gradient indicates an increase of electron velocity in the AGNR. When $\varepsilon= \varepsilon_2$, we obtain a linear dispersion curve, i.e. $|E| \propto k$ with zero band gap at k=0, similar to the massless dirac fermions in 2D-graphene. As $\varepsilon$ further increases, the band gap increases again, and achieves the maximum band gap value (SC-state) at $\varepsilon= \varepsilon_3$.

The E-field induced electronic transformations that occur in the multilayer AGNR are useful for electronic device applications. The change in band gap as well as the M/SC transition can be utilized to increase the ON/OFF ratio in the field-effect-transistor devices. By using the transversal E-field, together with conventional (vertical) E-field, we can dramatically improve the sub-threshold slope of the FETs. In a normal AGNR, the electron velocity is generally small compared to the graphene sheet. The increase in the velocity, obtained by transversal E-field, would further enhance the mobility of the ribbon, and thus enhance the performance of the electronic devices.

In summary we showed that the electronic properties of a multilayer AGNR can be modified by the application of transversal E-field. Various phenomena such as band gap modulation, M-SC transition, change in electron velocity and the sign of the electron effective mass, as well as formation of linear dispersion with massless Dirac fermions were noticed. These phenomena are useful for device application.

Acknowledgments

This work was supported by ONR, NSF, and ARL.

**Captions**

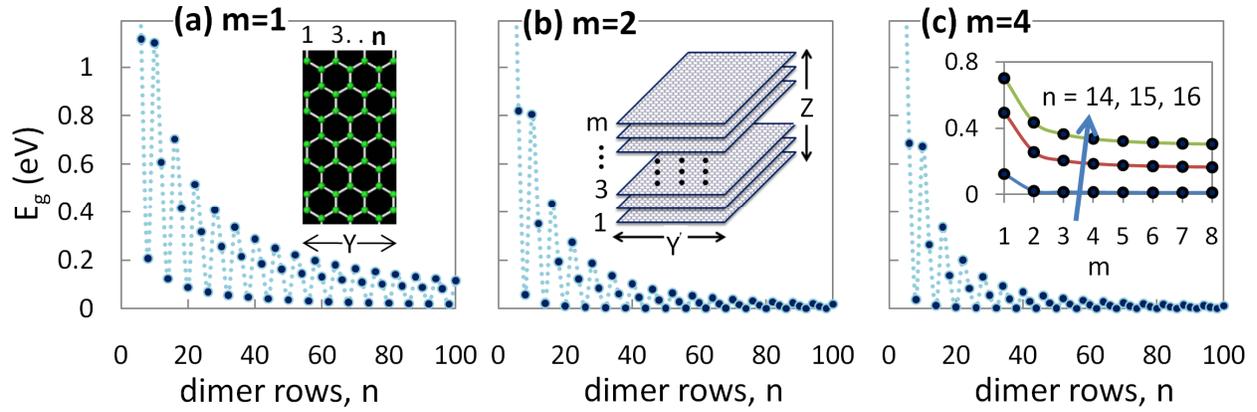

Figure 1 Band gap of multilayer AGNRs with increasing width when (a) number of layers, m=1 (b) m=2, and (c) m=4. In each case there are three distinct curves. The magnitude of the band gap is suppressed with increases m. The inset of (a) shows the structure of a single AGNR layer, (b) shows the structure of a multilayer AGNR, and (c) shows the band gap variation when m increases at fixed n. Note that n=14,15, and 16 represents (3p-1), (3p), and (3p+1) -AGNRs, respectively. Note that $Y = \sqrt{3}a_{cc}(n-1)/2$ and $Z = c(m-1)/2$.

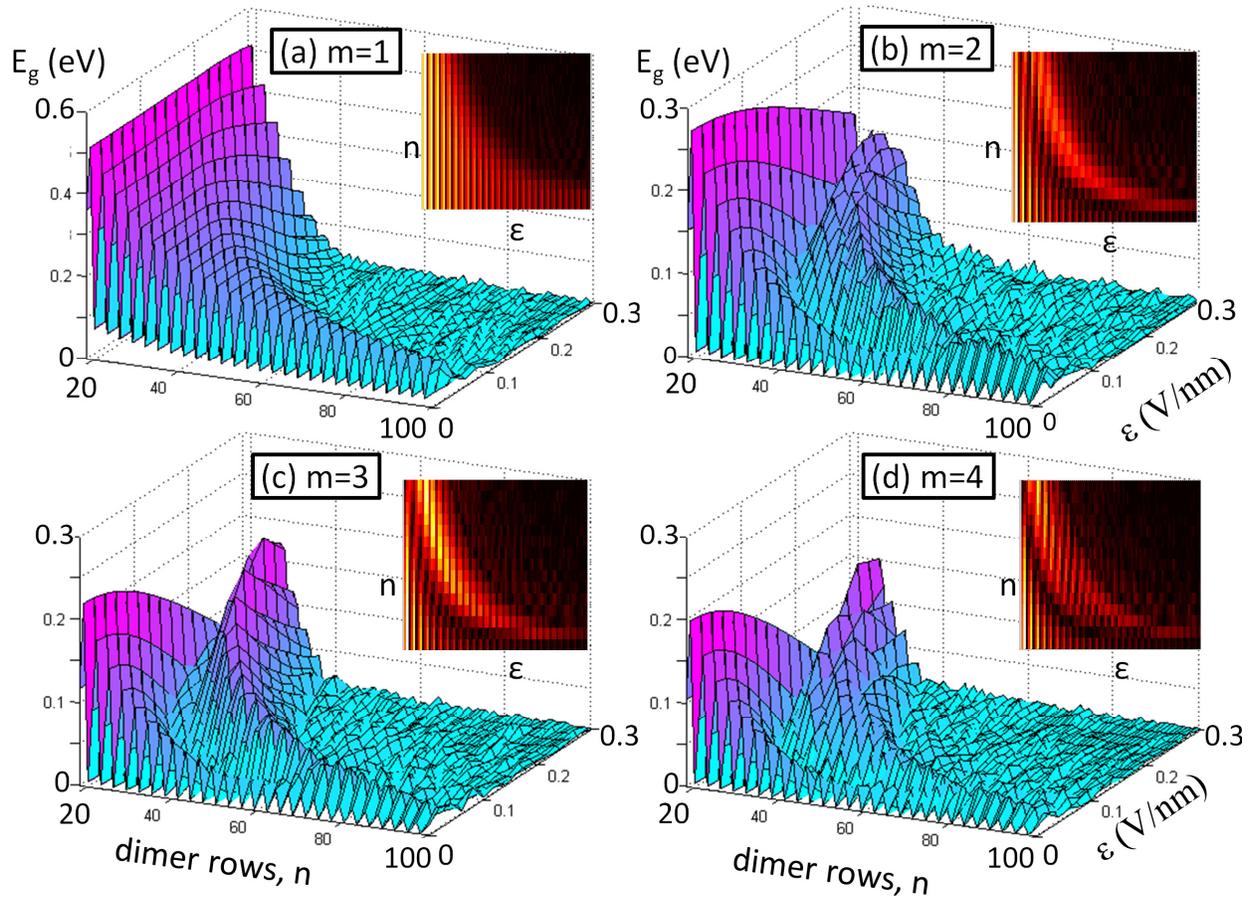

Figure 2 Band gap, $E_g$ of armchair GNR, with varying width and electric field ε for (a) number of layers, m=1, (b) m=2, (c) m=3, and (d) m=4. The inset in each figure shows the top view of the plots. For m=1, the $E_g$ decrease when ε increases. For multilayer AGNRs the $E_g$ increases with ε for optimized width.

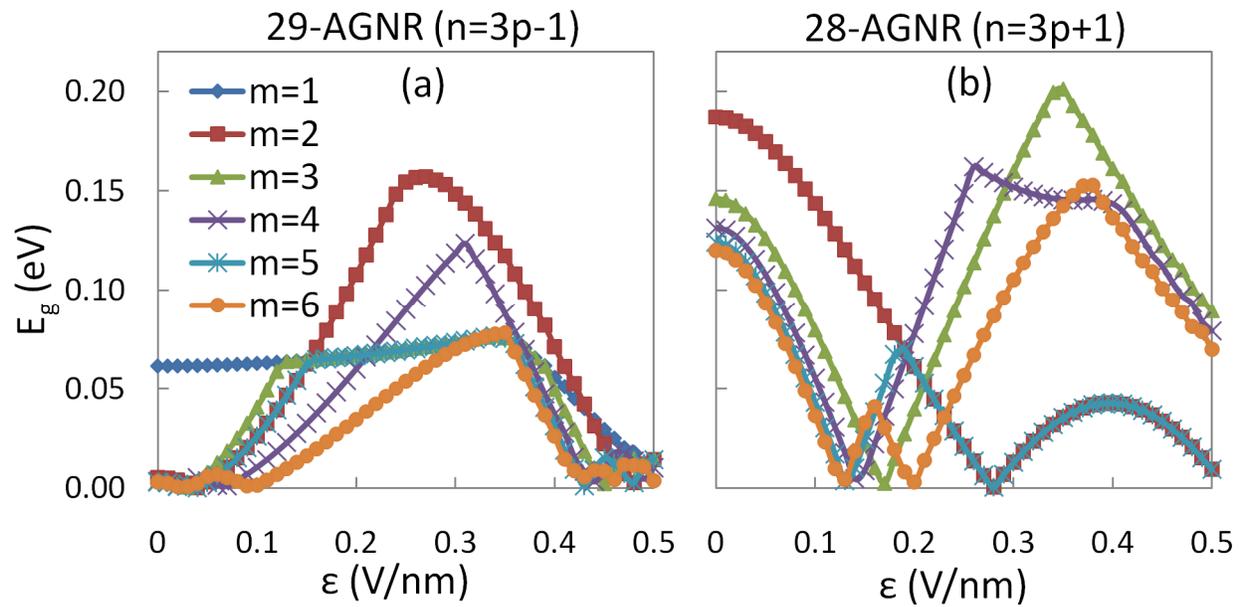

Figure 3 Bandgap variation with E-field, ε for (a) 29-AGNR (3p-1), and (b) 28-AGNR (3p+1) for different number of layers, m. For multilayer, i.e. m>1, 29-AGNR is metallic (M) and 28-AGNR is semiconducting (S) at ε=0. When the 29 (28)-AGNR changes from M-SC-M (SC-M-SC). A significant M-S transition is obtained when m=2 (m=3) for 29(28)-AGNR. Note that in (b), the plot for m=1 is not shown as the band gap values are larger than the values shown in the figure.

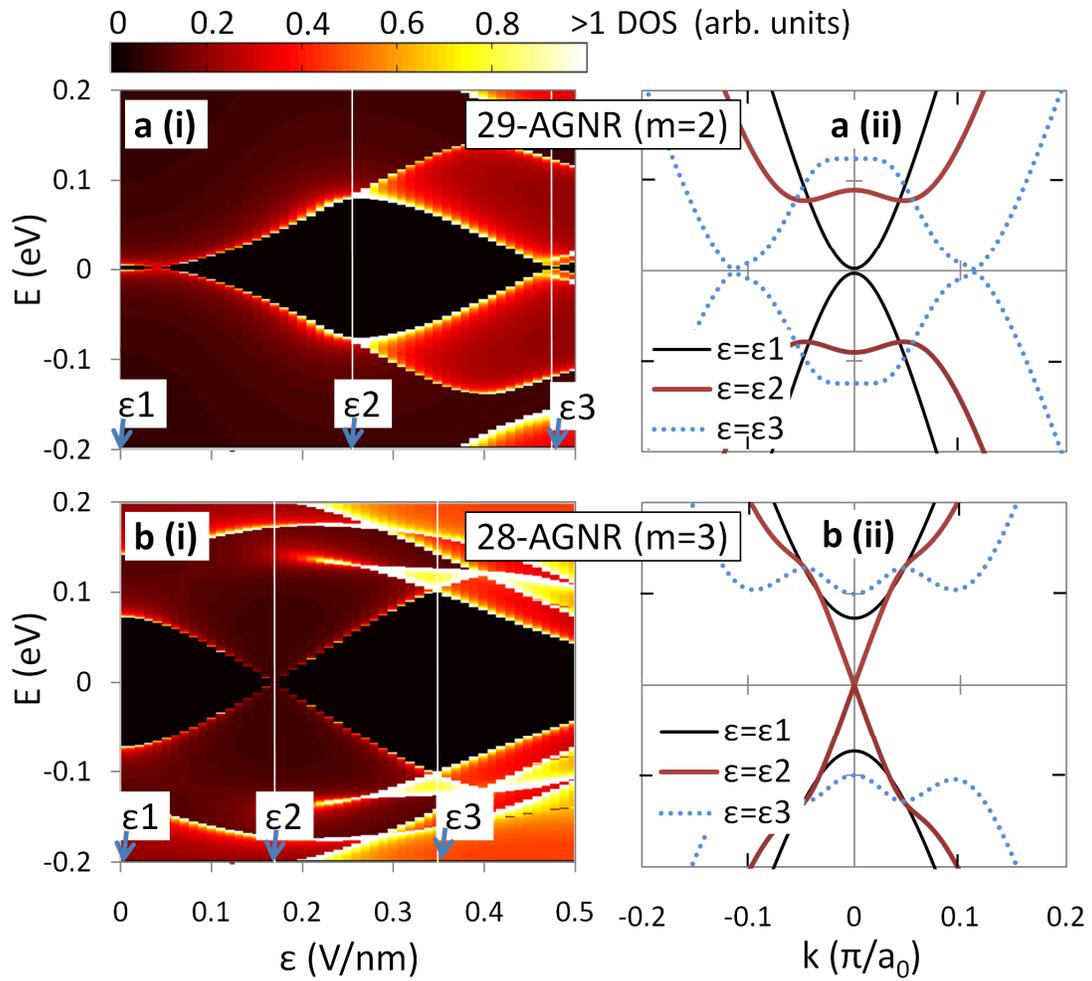

Figure 4 Density of state (DOS) map indicating (ai)M-SC-M transition for bilayer 29-AGNR, and (bi) SC-M-SC transition for trilayer 28-AGNR. (aii) shows bandstructure of bilayer 29-AGNR at ε= ε1, ε2, and ε3 as indicated in (ai). (bii) shows bandstructure of bilayer 29-AGNR at ε= ε1, ε2, and ε3 as indicated in (bi). Note that in (bi) and (bii) only the lowest (largest) conduction (valence) band is shown.